%% file: ksu_paper.tex
\documentclass[12pt]{article}
\pdfoutput=1
\usepackage{amsmath}
\usepackage{mathtools}
\usepackage{graphicx}
\usepackage{color}
\usepackage{natbib}
\usepackage{setspace}
\usepackage{amsfonts}
\usepackage{authblk}
\usepackage{url}

\usepackage[top=1.5in, bottom=1in, left=1in, right=1in]{geometry}

\newcommand{\I}{\mathrm{I}}

\bibpunct{(}{)}{;}{a}{}{,} 

\title{Bayesian Inference for a Covariance Matrix}
\author[1]{Ignacio Alvarez }
\author[1]{Jarad Niemi }
\author[2]{ Matt Simpson}

\affil[1]{Department of Statistics, Iowa State University}
\affil[2]{Department of Statistics and Department of Economics, Iowa State University}

\date{August 2014}
\begin{document}
 
\maketitle

\begin{abstract}
Covariance matrix estimation arises in multivariate problems including multivariate normal sampling models and regression models where random effects are jointly modeled, e.g. random-intercept, random-slope models. A Bayesian analysis of these problems requires a prior on the covariance matrix. Here we compare an inverse Wishart, scaled inverse Wishart, hierarchical inverse Wishart, and a separation strategy as possible priors for the covariance matrix. We evaluate these priors through a simulation study and application to a real data set. Generally all priors work well with the exception of the inverse Wishart when the true variance is small relative to prior mean. In this case, the posterior for the variance is biased toward larger values and the correlation is biased toward zero. This bias persists even for large sample sizes and therefore caution should be used when using the inverse Wishart prior.
\end{abstract}


\section{Introduction} 

Covariance matrix estimation arises in multivariate problems including multivariate normal sampling models and regression models where random effects are jointly modeled, e.g. random-intercept, random-slope models. Bayesian estimation of a covariance matrix requires a prior for the covariance matrix . The natural conjugate prior for the multivariate normal distribution is the inverse Wishart distribution \citep{barnard2000}. Due to its conjugacy, this is the most common prior implemented in Bayesian software. However, this prior has issues: the uncertainty for all variances is controlled by a single degree of freedom parameter \citep{bda2013}, the marginal distribution for the variances has low density in a region near zero \citep{gelman2006prior}, and there is an \emph{a priori} dependence between correlations and variances \citep{visualize}. These characteristics of the prior can impact posterior inferences about the covariance matrix. 

Alternative covariance matrix priors have been proposed including the scaled inverse Wishart \citep{odomain}, a hierarchial inverse Wishart \citep{huang2013simple}, and a separation strategy \citep{barnard2000}. However \cite{visualize} states that 
\begin{quote}
Even fewer analytical results are known for these families, making it even more challenging to understand precisely the properties of such distributions. Consequently, our analytical understanding of these distributions falls short of providing us a full understanding of the inverse-Wishart distribution.
 \end{quote}
 
The objective of this study is to understand the impact of these prior choices on the posterior inference of the covariance matrix. We select some of the proposed prior models in the literature and first run a simulation study to assess the impact on posterior, then we apply each model to a real data set consisting of bird counts in national forest in the Great Lakes. 

The rest of this paper is organized as follows: Section \ref{sec:models} describes the statistical methods and the covariance prior distributions we use. Section \ref{sec:results} presents a simulation study consisting in simulate data from a multivariate normal model and make inference about the covariance matrix to compare the different priors. Section \ref{sec:birds} presents an application to bird counts in the Superior National Forest. Finally, Section \ref{sec:summary} provides the applied modeler with strategies for estimating a covariance matrix. 

\section{Statistical Models \label{sec:models} }

Consider the multivariate normal model, that is let $Y_i\in \mathbb{R}^d$ for $i=1,\ldots,n$ and assume $Y_i \stackrel{iid}{\sim} N(\mu, \Sigma)$ with $\mu\in \mathbb{R}^d$ and $\Sigma$ is a $d$-dimensional positive definite matrix.   
  \begin{equation}
 p(y\vert \mu,\Sigma) \propto |\Sigma|^{-n/2} \mbox{exp}\left\{- \frac{1}{2} \sum_{i=1}^n (y_i-\mu)^\top \Sigma^{-1} (y_i-\mu) \right\} = |\Sigma|^{-n/2}  \mbox{exp}\left\{- \frac{1}{2}  \mbox{tr}(\Sigma^{-1}S_\mu)  \right\} 
 \label{like}
 \end{equation}
The likelihood is provided in equation \eqref{like} where $y$ represents the entire data and  $S_\mu = \sum_{i=1}^n (y_i-\mu) (y_i-\mu) ^\top$. 

For this study, the primary parameter of interest is the matrix $\Sigma$ with elements $\Sigma_{ij}$. We will often refer to the standard deviations $\sigma_i$ and correlations $\rho_{ij}$ where $\sigma_i^2 = \Sigma_{ii}$ and $\Sigma_{ij} = \rho_{ij}\sigma_i\sigma_j$. 

In the following subsections, we introduce a number of covariance matrix prior classes: inverse Wishart, scaled inverse Wishart, hierarchical inverse Wishart, and a separation strategy. 

\subsection{Inverse Wishart prior \label{sec:iw}}

The natural conjugate prior for a covariance matrix is the inverse Wishart (IW) prior \citep{barnard2000}. 
\begin{equation} 
p(\Sigma) \propto  |\Sigma|^{-(\nu+ d +1)/2 } e^{-\frac{1}{2} tr( \Lambda \Sigma^{-1}) }
\label{eq:wis}
\end{equation}
The IW prior has density provided in equation \eqref{eq:wis} where $\Lambda\in \mathbb{R}^d$ is a positive definite $d$ dimensional matrix and $\nu$ is a scalar degrees of freedom. For a proper prior, $\nu>d-1$. The mean is $E(\Sigma) = \Sigma_0= \frac{\Lambda}{\nu - d - 1}$ for $\nu>d+1$. 

The IW prior is commonly used due to its conjugacy properties with the normal sampling model. Specifically, the full conditional for $\Sigma$ is $\Sigma \vert y,\mu \sim IW(n+\nu_0, \Lambda_0+S_\mu)$. If $\mu|\Sigma \sim N(\mu_0,\Sigma/\kappa_0)$, then the marginal posterior for $\Sigma$, $\Sigma|y$, has an IW distribution (see Sec 3.6 of \cite{bda2013}). This conjugacy allows easy incorporation into Markov chain Monte Carlo (MCMC) approaches based on Gibbs sampling.

Using an inverse Wishart prior for the covariance matrix implies a scaled inverse chi-square distribution\footnote{The scaled inverse chi-square denoted by $X \sim \mbox{inv-}\chi^2(\nu, s^2)$ has a density function given by $p(x) =  \frac{(\nu/2)^{\nu/2}} {\Gamma(\nu/2)} s^{\nu}x^{-(\nu/2 + 1)} \mbox{exp}\left\{-\nu s^2 / 2x\right\} $} for each variance $\sigma_i^2\sim \mbox{inv-}\chi^2(\nu - d + 1, \frac{\lambda_{ii}}{\nu-d+1} )$ where $\lambda_{ii}$ is the $i^{th}$ diagonal entry of $\Lambda$. If $\Lambda$ is a diagonal matrix, each correlation has marginal density as $p(\rho_{ij}) \propto (1 - \rho_{ij}^2)^{(\nu - d + 1)/2}$. A default approach for the IW prior sets $\Lambda=\I$ and $\nu=d+1$ where $\I$ is an identity matrix. A diagonal $\Lambda$ and $\nu=d+1$ results in marginal uniform distributions for all correlations. 

There are a least three problems with the IW prior. First, the uncertainty for all variance parameters is controlled by the single degree of freedom parameter and thus provides no flexibility to incorporate different amounts of prior knowledge to different variance components (see Sec 19.2 of \cite{bda2003}). Second, when $\nu>1$, the implied scaled inv-$\chi^2$ distribution on each individual variance has extremely low density in a region near zero and thus causes bias in the result posteriors for these variances \citep{gelman2006prior}.  Third, the IW prior imposes a dependency between the correlations and the variances. In particular larger variances are associated with absolute values of the correlations near 1 while small variances are associated with correlations near zero \citep{visualize}.  We illustrate the latter two problems in Section \ref{sec:results}.


\subsection{Scaled Inverse Wishart \label{sec:siw}}

An alternative to the IW prior is the scaled inverse Wishart (SIW) prior which is based on the inverse Wishart distribution but adds additional parameters for flexibility \citep{odomain}. The SIW prior defines $\Sigma \equiv \Delta Q \Delta $ where $\Delta$ is a diagonal matrix with $\Delta_{ii}=\delta_i$, then 

\begin{equation}
Q \sim  IW(\nu, \Lambda) \;\;, \;\; \log(\delta_i) \stackrel{ind} \sim N(b_i, \xi_i^2)
\label{eq:siw}
\end{equation} 

We use the notation $\Sigma \sim SIW(\nu, \Lambda, b, \xi)$ to refer to this prior.  By construction, the SIW prior implies that $\sigma_i = \delta_i \sqrt{Q_{ii}}$, and $\Sigma_{ij}=\delta_i\delta_jQ_{ij}$. Thus each standard deviation is the product of a log-normal and the square root of a scaled inv-$\chi^2$ and the correlations $\rho_{ij} = Q_{ij}/\sqrt{Q_{ii}Q_{jj}}$ have the same distribution they had under the inverse Wishart on $Q$.  


This prior is recommended by \cite{gelmanhill}, setting $\nu=d+1$ and $\Lambda=\I$ to ensure uniform priors on the correlations as in the IW prior but now there is more flexibility on incorporating prior information about the standard deviations. 

\subsection{Hierarchical Half-t prior}

Recently, \cite{huang2013simple} proposed a hierarchical approach for the covariance matrix as shown in equation \eqref{eq:ht}. 

\begin{equation}
\Sigma \sim IW( \nu + d - 1 ,  2\nu\Lambda) \;\;,\;\;  \lambda_i  \stackrel{ind} \sim \mbox{Ga}\left(\frac{1}{2} , \frac{1}{\xi_i^2}\right) \;\; \mbox{with} \; E(\lambda_i)=\frac{\xi_i^2}{2} 
\label{eq:ht}
\end{equation} 
where $\Lambda$ is diagonal matrix with $i^{th}$ element $\lambda_i$. 

This prior implies a half $t$ distribution with $\nu$ degrees of freedom, location parameter 0, and scale parameter $\xi_i$.  The marginal distribution implied for correlations is giving by $p(\rho) \propto (1-\rho^2)^{\frac{\nu_0}{2}-1}$ then letting $\nu=2$ implies marginally uniform distribution for the correlation coefficient. We use the notation $\Sigma \sim HIW_{ht}(\nu, \xi)$ to denote this prior.

A similar approach was proposed by \cite{daniels1999} and \cite{matilde}, they use flat priors for the diagonal entries of $\Lambda$ matrix and put a prior on the degrees of freedom parameter.  An important advantage of  HIW$_{ht}$ is the implied half-t prior on the standard deviations which is recommended by \cite{gelman2006prior}.

\subsection{Separation Strategy \label{sec:ss} }

\cite{barnard2000} propose the a separation strategy (SS) where the standard deviations and correlations are modeled independently and then combined to form a prior on the covariance matrix. They decompose the covariance as $\Sigma = \Lambda R \Lambda$  where $\Lambda$ is a diagonal matrix with the $i^{th}$ element $\sigma_{i}$ and $R$ is a correlation matrix with $\rho_{ij}$ as the $i^{th}$ row and $j^{th}$ column element of $R$. 

\cite{barnard2000} constructs a correlation matrix $R$ from an inverse Wishart distribution. Specifically, let $Q\sim IW(\nu, I )$ then $R = \Delta Q \Delta$ where $\Delta$ is a diagonal matrix with $i^{th}$ diagonal element $Q_{ii}^{-1/2}$. The prior density for the correlation matrix is $p(R) \propto |R|^{-\frac{1}{2}(\nu+k+1) }  (\prod_{i=1}^k r^{ii}) ^{\frac{\nu}{2}}$, where $r^{ii}$ is the $i$th diagonal element of $R^{-1}$. They then assume a log-normal distribution for the logarithm of the standard deviations, i.e. $\log(\sigma_i) \stackrel{iid} \sim N(b_i, \xi_i)$.  We use the notation $\Sigma \sim \mbox{BMM}_{mu}(\nu,b,\xi)$ to denote this prior.

Under the BMM$_{mu}$ specification, individual correlations have the marginal density $p(\rho_{ij}) \propto (1-\rho_{ij}^2)^{\frac{\nu-d-1}{2}}$ and then setting $\nu=d+1$ lead to marginally uniformly distributed correlations which are also independent from the variances by construction. 

The main disadvantage of BMM$_{mu}$ is computational. 
With more recently developed software, e.g. Stan \citep{stan2014}, this restriction is not as  detrimental since this software is not based on Gibbs sampling but instead on Hamiltonian Monte Carlo (HMC) which is a Metropolis strategy for all parameters simultaneously.  In fact, the STAN manual \citep{stanmanual2014} recommends following a separation strategy, but suggests the LKJ prior for the correlation \cite{lewandowski2009generating}. In high dimensions ($d>10$), the LKJ prior concentrates mass near zero correlations and therefore we do not use it here.

\section{Simulation study results \label{sec:results}}

In this section we carry out a simulation based analysis to assess the performance of these different covariance matrix prior. First, we simulate from each prior to study the \emph{a priori} relationship between correlations and standard deviations. Second, we simulate data from the model and analyze posterior means to determine the impact prior choice has on posterior inference. Table \ref{paramvals} presents the hyperparameter values used in both simulations study for every prior. 

We started with the default IW$(d+1,\I)$ prior which implies a marginal uniform prior on the correlations and an IG$(1,1/2)$ marginal prior for the variances. We set hyperparameters for the other three priors to imply marginal uniform priors for the correlations and prior medians on the variances to match the median of an IG$(1,1/2)$ of 0.72.    

\begin{table}[htbp]
   \centering
    \caption{ Parameter values for simulation study}
   \label{paramvals} 
   \begin{tabular}{ l|c|c}
   \hline
      Prior    &  Values for prior sampling & Values for posterior inference\\ \hline
  IW$(\nu, \Lambda)$ &   $\nu=d+1$, $\Lambda=\I$  & $\nu=d+1$, $\Lambda=\I$  \\ 
  SIW$(\nu, \Lambda, b, \xi)$  & $b=0$, $\xi_i =1$,  $\nu_0= d + 1$, $\Lambda = 0.8\I$ & $b=0$, $\xi_i =100$,  $\nu_0= d + 1$, $\Lambda = \I$ \\
  HIW$_{ht}(\nu, \xi)$    &  $\nu=2$,  $\xi_i=1.04$ & $\nu=2$,  $\xi_i= \sqrt{1000}$ \\
   BMM$_{mu}(\nu,,b,\xi)$   &  $\nu=d+1$, $b_i=\log(.72)/2$ , $\xi_i=1$ &  $\nu=d+1$, $b_i=0$ , $\xi_i=100$ \\ \hline
   \end{tabular}
 \end{table}

\subsection{Sampling from the priors} 

\begin{figure}[htbp]
\begin{center}
 \includegraphics[width=\textwidth ]{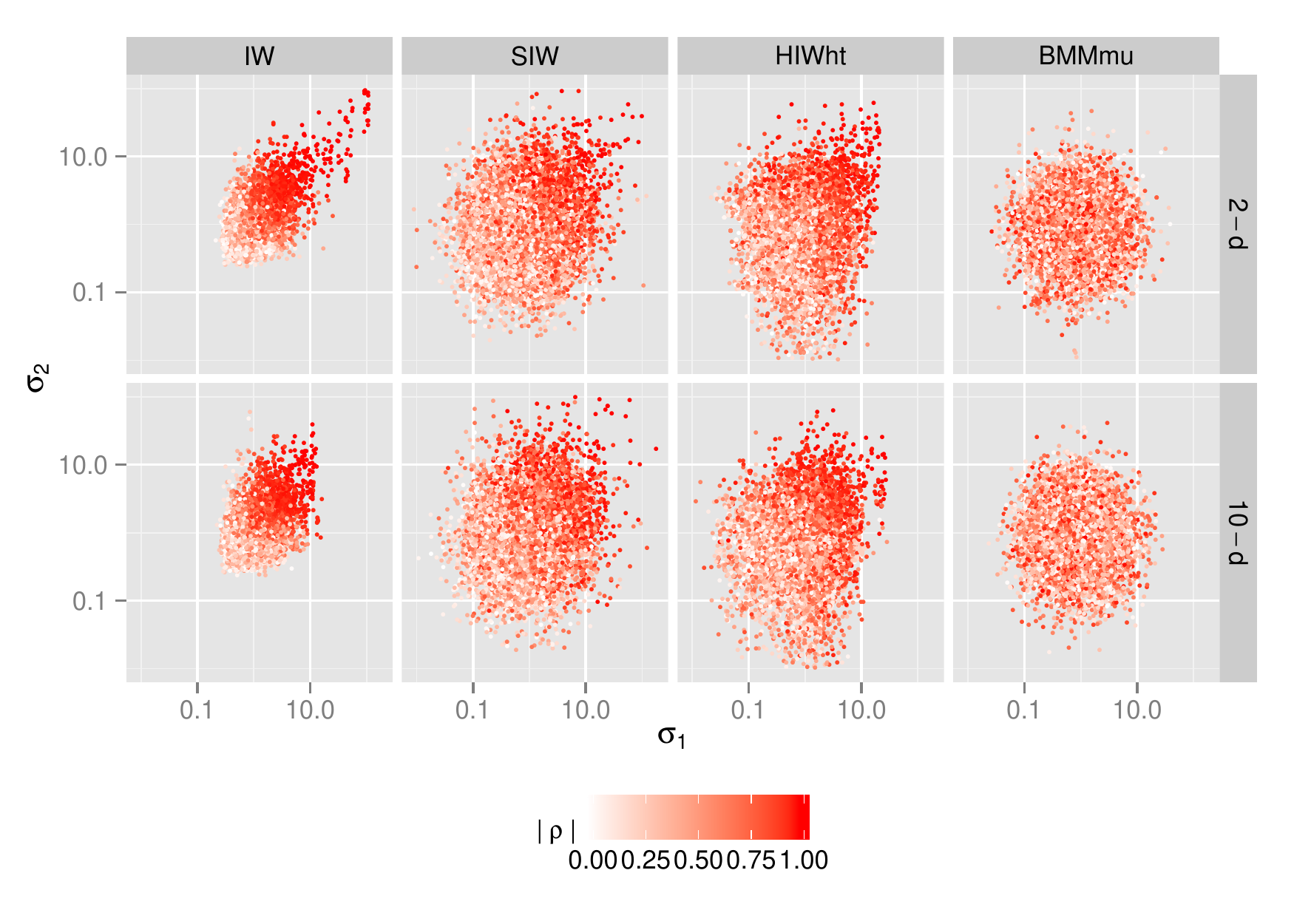} 
  \vspace{-.5in}
\caption{Samples from the prior distribution for the first two standard deviations coded by absolute value of the correlation for the four covariance matrix priors (columns) for a two- and ten-dimensional matrix (rows).}
\label{priorF2} 
\end{center}
\end{figure}

Figure \ref{priorF2} shows a scatterplot with the first two standard deviations from 10,000 draws from each of the four priors. The IW prior implies a positive relationship among the standard deviations. Also large correlations values appear more predominantly when the two variances are high, while low correlations appears when variances are low. 

The SIW and HIW$_{ht}$ priors show a similar picture to the IW prior but the dependence is weaker. For instance, the high correlations are mostly present when standard deviations are also high, however we do see some red points for small values of the standard deviations. In contrast, the BMM$_{mu}$ prior displays draws that are independent by construction.

\subsection{Impact on posterior} 

The prior relationship between standard deviations and correlations presented in the previous section may or may not be an issue depending on how much this relationship impacts the posterior. To concentrate on estimation of the covariance matrix and its constitutive components, we simulate mean-zero, normally distributed data in two and ten dimensions with equal variances in all dimensions. A similar approach is taken by  \cite{daniels1999} and \cite{matilde} in their simulations.

We study the sensitivity of correlation and standard deviation point estimates to the choice of the covariance matrix prior. Table \ref{scen} shows the sample size, standard deviation, and correlations used in the simulation study. We conducted a full factorial experiment with 5 replicates at each combination.


\begin{table}[htbp]
   \centering
   \caption{Simulation scenarios. Specific values used in simulations for each parameter. \label{scen}} 
     \begin{tabular}{lcc} \hline
          &  Bivariate    & Ten-dimensional  \\ \hline
      Sample size   ($n$)   & 10, 50, 250   &  10, 50  \\
      Standard deviation ($\sigma$)  & 0.01, 0.1, 1, 10, 100 & 0.1, 1, 100 \\
      Correlation ($\rho$)   &  0, 0,25, 0.5, 0.75, 0.99  &  0, 0.99 \\ \hline
   \end{tabular}
\end{table}

To reduce Monte Carlo variability in the simulations, data were simulated for $\sigma=1$ and then rescaled to get obtain simulations with the other variances. For instance, five replicate data sets that consists of ten observations from a bivariate independent standard normal, i.e. $n=10$, $d=2$ $\rho=0$ and $\sigma=1$, were simulated. Each simulated value is multiplied by 0.01 (or 0.1, 10, and 100) to obtain data sets with standard deviation approximately 0.01 (or 0.1, 10, and 100).

The models are estimated using Stan software (\cite{stan2014}) which uses a HMC algorithm to construct posterior samples with the No U-turn sampler (NUTS, \cite{hoffman2011no}) strategy. Default initial values were used. Model convergence is monitored using the potential scale reduction factor \citep{bda2003}.
Initially, 3 chains with 1000 iterations after burn-in is used for every model. Whenever the reduction factor is bigger than 1.1 a longer model with 2000 iterations per chain after burn-in period is estimated.  

\subsubsection{Inference for Correlation coefficient}

For inference, the true sampling model is assumed and the covariance matrix priors shown in Table \ref{paramvals} were assumed with the parameter values for posterior inference column. Only the bivariate results are shown here but the ten-dimensional results are similar.

\begin{figure}[hbtp]
\centering
\includegraphics[width=\textwidth] {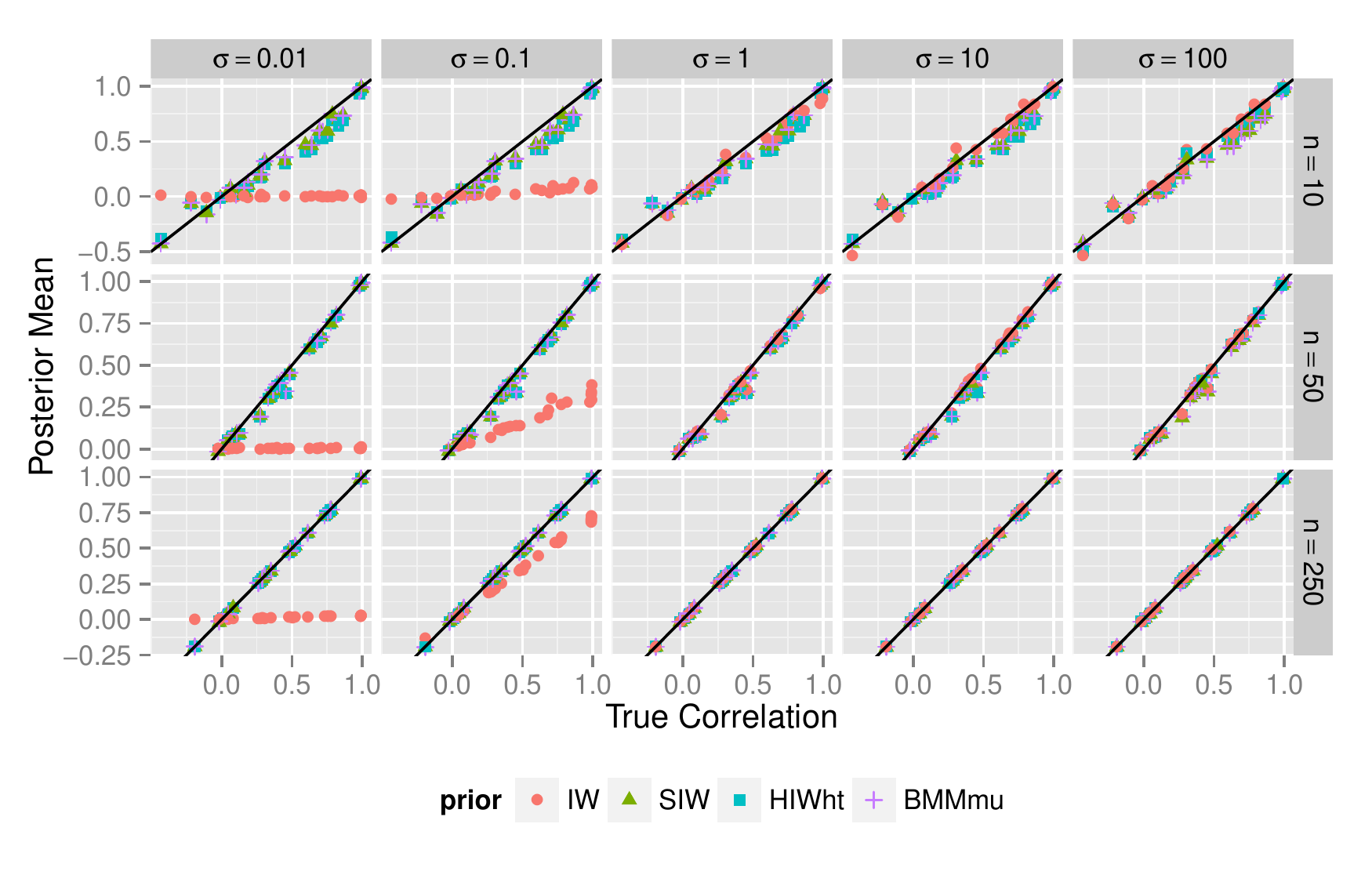} 
\vspace{-.5in}
\caption{Posterior mean of the correlation (vertical axis) versus jittered true correlations for the range of true standard deviations (columns) and sample sizes (rows) using the four covariance matrix priors.}
\label{rhod2}
\end{figure}

Figure \ref{rhod2} shows the estimated posterior means compared with the values used in the simulation. This figure shows that when the standard deviation is small the IW prior heavily shrinks the posterior correlation towards 0 even if the true correlation is close to 1. This bias is attenuated as the sample size increases, but is still remarkably large with 250 observations when the standard deviation is around 0.01. 

All other priors appear to recover the true correlation with a slight bias toward zero for small sample sizes.

\begin{figure}[htbp]
\centering
\includegraphics[width=\textwidth]{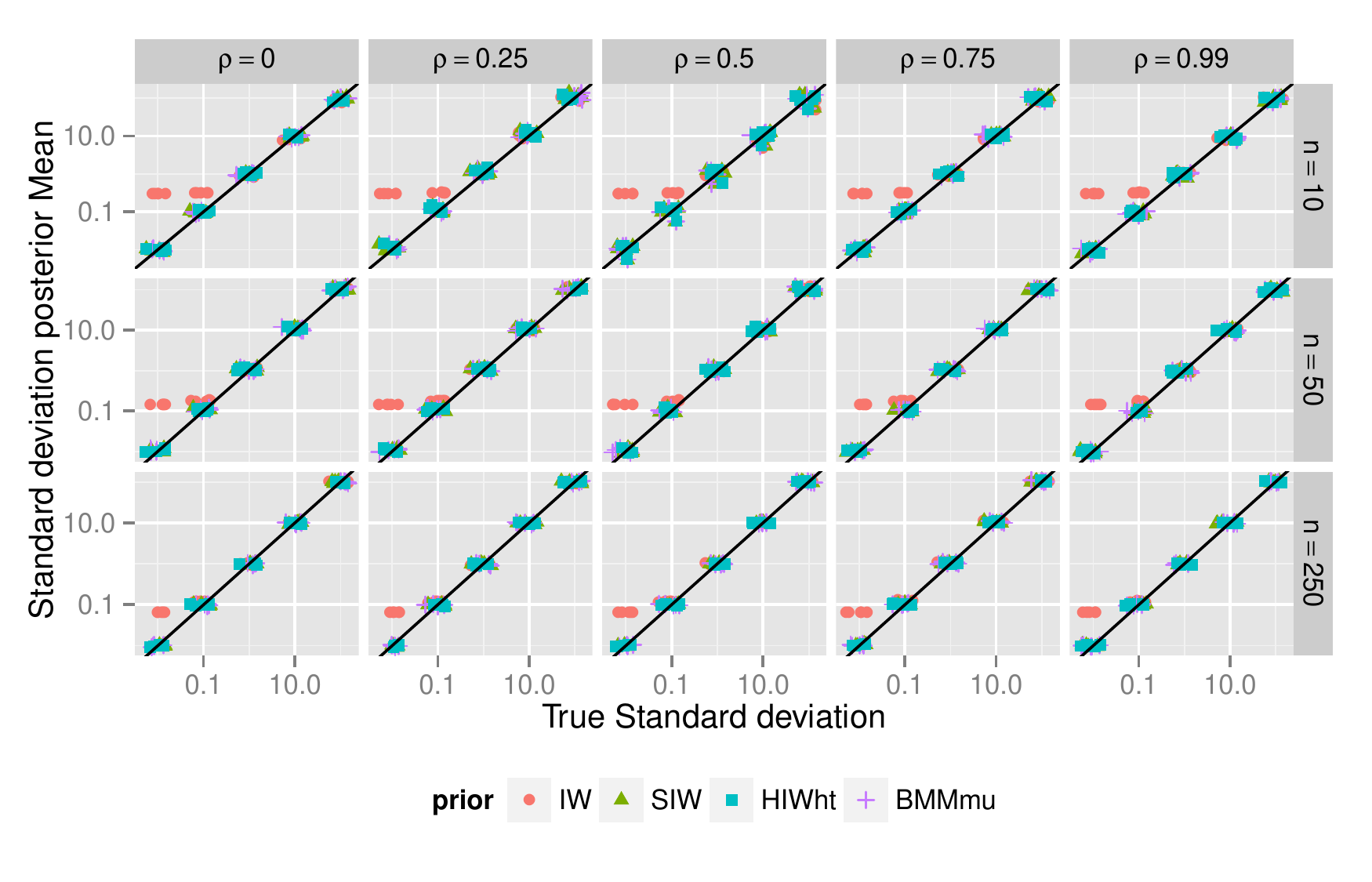} 
\vspace{-.5in}
\caption{Posterior mean of the first standard deviation (vertical axis) versus jittered true standard deviations for the range of true correlations (columns) and sample sizes (rows) using the four covariance matrix priors.}
\label{devF1} 
\end{figure}

Figure \ref{devF1} shows a scatterplot of the standard deviation posterior mean against the values used in the simulation. The figure shows that the IW prior overestimates the standard deviation for true values of the standard deviation that are small.  This bias is attenuated somewhat as sample size increase, but when the true standard deviation is 0.01 the bias is still quite large even with 250 observations. 

All other priors appear to have no difficulty in accurately estimating the standard deviation.  


We also assessed the posterior means for the covariance, the product of the correlation and two standard deviations, and found no evidence of bias for any of the priors \citep{CCalvarez}. For the IW prior, the lack of bias here appears to be due to the marginal prior for the covariance, which does not have a closed form, having heavy tails. 

Under these simulation conditions, we find the IW prior to have no bias in estimating the covariance, to be biased toward large values for small variances, and to be biased toward zero for correlations when the variance is small. Intuitively, the heavy tailed-ness in the marginal prior for the covariance results in robust estimation for the covariance, but the lack of prior mass near zero in the marginal prior for the variance results in an upward bias. Since a correlation is a covariance divided by two standard deviations, the only way to reconcile these results is to have a corresponding bias in the correlation toward zero.

\section{Bird counts in the Superior National Forest \label{sec:birds}}

The Natural Resources Research Institute conducts a long-term monitoring program in Superior National Forest to track regional avian population trends. The specific data set which it is used in this manuscript consist of the yearly bird counts for the 10 most abundant species from 1995 to 2013. 

Table \ref{tab:bird} provides summary statistics for the these data where total count is the actual number of birds counted within a year and mean count is the total count divided by the number of surveys, approximately 500 each year. The table then displays, for each species, the average and standard deviation of these total counts and mean counts across the years.

\input{bird_stat}

From a scientific perspective, the choice of whether to use the total count versus the mean count appears arbitrary when the purpose is to understand the correlation amongst species. But, due to our simulation results, we expect here that if we choose total count the correlation amongst the species will be well estimated. In contrast, if we choose mean count, the correlation will be biased toward zero and the estimated standard deviation will be biased toward larger values when using the IW prior.

\subsection{Correlation among Bird Species}

We run analyses of these data assuming a multivariate normal for each year with a mean of zero and an unknown covariance matrix. We assume the four prior distributions described in Table \ref{paramvals} as well as the IW prior using the scaled data. We also perform the analysis for each pair of species independently using a 2-dimensional normal and also simultaneously for all species using a 10-dimensional multivariate normal. Finally, we consider using the seemingly abitrary choice of total count and average count. 

These models are estimated with same methods used for the models with simulated data, using Stan software \citep{stan2014}. 

\begin{figure}[hbpt]
\centering
\includegraphics[width=\textwidth]{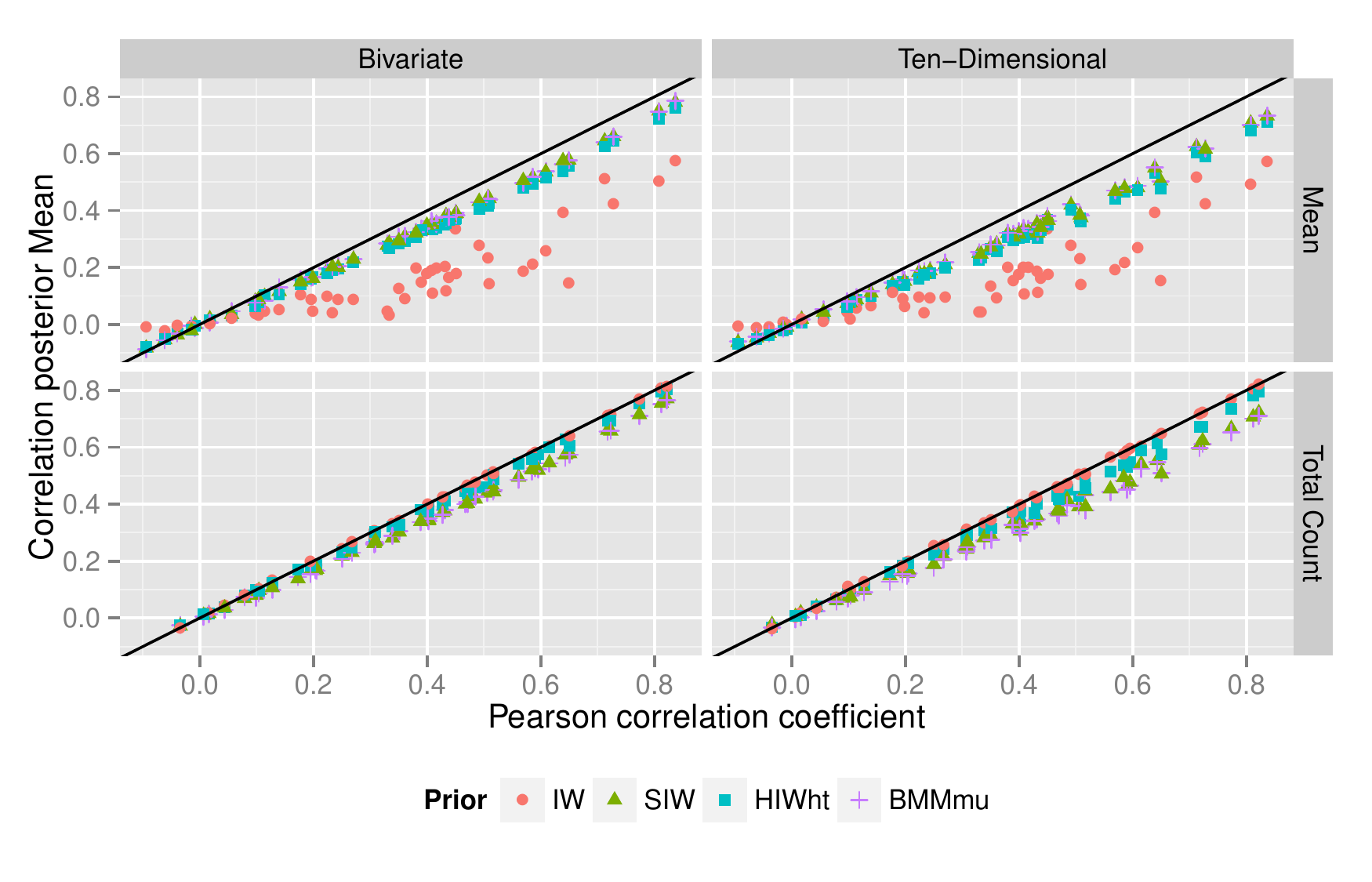}
 \vspace{-.5in}
\caption{Posterior mean of the correlation for total count (bottom row) versus average count (top row) when performing the analysis for each pair of species independently (left column) versus all species simultaneously (right column) for four different priors on the covariance matrix as well as the IW prior using scaled data (IWsc).}
\label{fig:coring}
\end{figure}
%

Figure \ref{fig:coring} presents the estimated correlations. The posterior mean for each model is plotted against the Pearson correlation coefficient. The four panels represent each combination of response and dimension and within each panel there are four estimates for each pair of bird species corresponding to the five priors used. 

These results match those found when using simulated data. When we model mean count, we see each correlation estimate is shrunk towards zero, but the IW shrinkage is extreme. However the results change if we decide to model total count. Now there is no shrinkage towards small correlation values and, in fact, the IW prior may be the best at recovering the Pearson correlation coefficient.

The SIW, HIW$_{ht}$ and BMM$_{mu}$ priors show similar behavior no matter which response is used to estimate the covariance matrix, and estimating correlation with any of these priors will lead to basically the same conclusions.

\section{Summary \label{sec:summary}} 

Simulations from the IW prior show a strong \emph{a priori} dependence between the correlation and the variances. The SIW and HIW$_{ht}$ priors show a similar dependence but both seem to be more flexible than the IW prior. The BMM$_{mu}$ prior presents the most flexibility since variances and correlations are independent by construction.

Posterior results for the IW prior show an extreme bias toward larger values for small empirical variances and a corresponding shrinkage of the estimated correlation toward zero. In contrast, the other three priors show only a slight bias toward zero correlations and this is attenuated completely for larger sample sizes.

From a modeling perspective, the BMM$_{mu}$ prior flexibility is appealing since we can model correlations and variances independently and let the data define their relationship.  The BMM$_{mu}$ prior is the original proposal of \cite{barnard2000}, however a separation strategy could use a different prior for variances and correlations, e.g. a half-t for the standard deviations and uniform for the correlations. 

From a computational perspective we expect the BMM$_{mu}$ prior to be the most complex, since all the others conserve the conjugacy properties of the IW distribution. This is especially relevant for a Gibbs sampler, where conjugacy allows for a Gibbs steps. Using Stan or any other HMC-based approach, there is relatively little cost to using other prior such as BMM$_{mu}$. 
Computationally, the SIW prior is more efficient than HIW or BMM even when using Stan \citep{CCalvarez}. 

In summary, the prior choice may depend on which are the computational resources are available. If it is possible to use a HMC sampler such as STAN, the separation strategy proposed by \cite{barnard2000} gives modeling flexibility and good inference properties. Whenever we use Gibbs based samplers, e.g. JAGS or BUGS, a prior which maintains some conjugacy might be preferable. For some Bayesian software, the IW prior may be our only option. In this case, if our only objective is to estimate the correlation, it is possible to pre-scale that data such that each component as an empirical variance of 1 and then proceed with the IW prior. This appears to provide unbiased estimates of the correlations \citep{CCalvarez}. 

\bibliographystyle{asa}      
\bibliography{report_year}      
\end{document}

%% file: bird_stat.tex
\begin{table}[ht]
\centering
\caption{Summary statistics of bird count data in Superior National Forest from
1995 to 2013 for the 10 most abundant species.} 
\label{tab:bird}
\begin{tabular}{lcccc}
  \hline
  \multicolumn{1}{c}{Species} & \multicolumn{2}{c}{Total count} & \multicolumn{2}{c}{Mean count} \\
Name & Average & Stand. Dev. &Average & Stand. Dev.  \\ 
  \hline
Ovenbird & 1098 & 244 & 2.18 & 0.51 \\ 
  White-throated Sparrow & 725 & 135 & 1.43 & 0.24 \\ 
  Nashville Warbler & 722 & 214 & 1.42 & 0.36 \\ 
  Red-eyed Vireo & 672 & 145 & 1.34 & 0.33 \\ 
  Chestnut-sided Warbler & 432 & 133 & 0.86 & 0.28 \\ 
  Veery & 293 & 65 & 0.58 & 0.13 \\ 
  Blue Jay & 267 & 85 & 0.52 & 0.15 \\ 
  American Robin & 225 & 84 & 0.44 & 0.15 \\ 
  Hermit Thrush & 222 & 67 & 0.44 & 0.13 \\ 
  Least Flycatcher & 136 & 35 & 0.27 & 0.06 \\ 
   \hline
\end{tabular}
\end{table}

%% file: ksu_paper.bbl
\begin{thebibliography}{15}
\newcommand{\enquote}[1]{``#1''}
\expandafter\ifx\csname natexlab\endcsname\relax\def\natexlab#1{#1}\fi

\bibitem[{Alvarez(2014)}]{CCalvarez}
Alvarez, I. (2014), \enquote{Bayesian Inference for a covariance matrix,}
  Master's thesis, Iowa State University, \url{http://arxiv.org/abs/1408.4050}.

\bibitem[{Barnard et~al.(2000)Barnard, McCulloch, and Meng}]{barnard2000}
Barnard, J., McCulloch, R., and Meng, X.-L. (2000), \enquote{Modeling
  covariance matrices in terms of standard deviations and correlations, with
  application to shrinkage,} \textit{Statistica Sinica}, 10, 1281--1312.

\bibitem[{Bouriga and F{\'e}ron(2013)}]{matilde}
Bouriga, M. and F{\'e}ron, O. (2013), \enquote{Estimation of covariance
  matrices based on hierarchical inverse-Wishart priors,} \textit{Journal of
  Statistical Planning and Inference}, 143, 795--808.

\bibitem[{Daniels and Kass(1999)}]{daniels1999}
Daniels, M.~J. and Kass, R.~E. (1999), \enquote{Nonconjugate Bayesian
  estimation of covariance matrices and its use in hierarchical models,}
  \textit{Journal of the American Statistical Association}, 94, 1254--1263.

\bibitem[{Gelman(2006)}]{gelman2006prior}
Gelman, A. (2006), \enquote{Prior distributions for variance parameters in
  hierarchical models,} \textit{Bayesian Analysis}, 1, 515--533.

\bibitem[{Gelman et~al.(2013)Gelman, Carlin, Stern, Dunson, Vehtari, and
  Rubin}]{bda2013}
Gelman, A., Carlin, J.~B., Stern, H.~S., Dunson, D.~B., Vehtari, A., and Rubin,
  D.~B. (2013), \textit{Bayesian data analysis}, CRC press.

\bibitem[{Gelman et~al.(2003)Gelman, Carlin, Stern, and Rubin}]{bda2003}
Gelman, A., Carlin, J.~B., Stern, H.~S., and Rubin, D.~B. (2003),
  \textit{Bayesian data analysis}, Chapman and Hall.

\bibitem[{Gelman and Hill(2007)}]{gelmanhill}
Gelman, A. and Hill, J. (2007), \textit{Data analysis using regression and
  multilevel/hierarchical models}, Cambridge University Press.

\bibitem[{Hoffman and Gelman(2011)}]{hoffman2011no}
Hoffman, M.~D. and Gelman, A. (2011), \enquote{The no-U-turn sampler:
  Adaptively setting path lengths in Hamiltonian Monte Carlo,} \textit{arXiv
  preprint arXiv:1111.4246}.

\bibitem[{Huang et~al.(2013)Huang, Wand, et~al.}]{huang2013simple}
Huang, A., Wand, M., et~al. (2013), \enquote{Simple marginally noninformative
  prior distributions for covariance matrices,} \textit{Bayesian Analysis}, 8,
  439--452.

\bibitem[{Lewandowski et~al.(2009)Lewandowski, Kurowicka, and
  Joe}]{lewandowski2009generating}
Lewandowski, D., Kurowicka, D., and Joe, H. (2009), \enquote{Generating random
  correlation matrices based on vines and extended onion method,}
  \textit{Journal of multivariate analysis}, 100, 1989--2001.

\bibitem[{O'Malley and Zaslavsky(2005)}]{odomain}
O'Malley, A.~J. and Zaslavsky, A.~M. (2005), \enquote{Domain-Level Covariance
  Analysis for Survey Data with Structured Nonresponse,}
  \url{http://www.stat.columbia.edu/~gelman/stuff_for_blog/omalley.pdf}.

\bibitem[{{Stan Development Team}(2014{\natexlab{a}})}]{stan2014}
{Stan Development Team} (2014{\natexlab{a}}), \enquote{Stan: A C++ Library for
  Probability and Sampling, Version 2.2,} .

\bibitem[{{Stan Development Team}(2014{\natexlab{b}})}]{stanmanual2014}
--- (2014{\natexlab{b}}), \textit{Stan Modeling Language Users Guide and
  Reference Manual, Version 2.2}.

\bibitem[{Tokuda et~al.(2011)Tokuda, Goodrich, Van~Mechelen, and
  Gelman}]{visualize}
Tokuda, T., Goodrich, B., Van~Mechelen, I., and Gelman, A. (2011),
  \enquote{Visualizing Distributions of Covariance Matrices,}
  \url{http://www.stat.columbia.edu/~gelman/research/unpublished/Visualization.pdf}.

\end{thebibliography}
